\documentclass[conference]{IEEEtran}
\IEEEoverridecommandlockouts
\usepackage{cite}
\usepackage{amsmath,amssymb,amsfonts}
\usepackage{algorithm}
\usepackage{algorithmic}
\usepackage{graphicx}
\usepackage{textcomp}
\usepackage{xcolor}
\usepackage{array}
\def\BibTeX{{\rm B\kern-.05em{\sc i\kern-.025em b}\kern-.08em
    T\kern-.1667em\lower.7ex\hbox{E}\kern-.125emX}}
\graphicspath{ {./Figures/} }
\begin{document}

\title{WirePlanner: Fast, Secure and Cost-Efficient Route Configuration for SD-WAN\\
}

\author{
\IEEEauthorblockN{Yunxi Shen}
\IEEEauthorblockA{\textit{Tsinghua University}}
\and
\IEEEauthorblockN{Yongxiang Wu}
\IEEEauthorblockA{\textit{Tsinghua University}}
\and
\IEEEauthorblockN{Junxian Shen}
\IEEEauthorblockA{\textit{Tsinghua University}}
\and
\IEEEauthorblockN{Han Zhang}
\IEEEauthorblockA{\textit{Tsinghua University}}
}

\maketitle

\begin{abstract}
As enterprises increasingly migrate their applications to the cloud, the demand for secure and cost-effective Wide Area Networking (WAN) solutions for data transmission between branches and data centers grows. Among these solutions, Software-Defined Wide Area Networking (SD-WAN) has emerged as a promising approach. However, existing SD-WAN implementations largely rely on IPSec tunnels for data encryption between edge routers, resulting in drawbacks such as extended setup times and limited throughput. Additionally, the SD-WAN control plane rarely takes both latency and monetary cost into consideration when determining routes between nodes, resulting in unsatisfactory Quality of Service (QoS).  We propose WirePlanner, an SD-WAN solution that employs a novel algorithm for path discovery, optimizing both latency and cost, and configures WireGuard tunnels for secure and efficient data transmission. WirePlanner considers two payment methods: Pay-As-You-Go, where users pay for a fixed amount of bandwidth over a certain duration, and Pay-For-Data-Transfer, where users pay for the volume of transmitted data. Given an underlay topology of edge routers and a user-defined budget constraint, WirePlanner identifies a path between nodes that minimizes latency and remains within the budget, while utilizing WireGuard for secure data transmission.
\end{abstract}


\section{Introduction}
In the modern digital era, the trend of moving enterprise applications to cloud-based platforms is on the rise. This shift necessitates the need for Wide Area Networking (WAN) solutions that are not only secure and dependable but also cost-effective to support data transfers across various locations. At the forefront of these solutions is Software-Defined Wide Area Networking (SD-WAN), which harnesses the concepts of software-defined networking (SDN) to offer enhanced adaptability and scalability. However, SD-WAN in its current form encounters notable obstacles, especially in the realms of ensuring security, reducing latency, and optimizing costs.

One of the primary concerns in SD-WAN deployments is ensuring the secure transmission of data between edge routers\cite{b1}. Most existing solutions rely on Internet Protocol Security (IPSec) tunnels for encryption\cite{b14, b15, b16}, which can introduce drawbacks such as long setup times and limited throughput. This can negatively impact the Quality of Service (QoS) experienced by users, hindering the adoption of SD-WAN in various enterprise contexts.

Another challenge in SD-WAN implementation is the control plane's ability to optimize both latency and monetary cost when finding a path between nodes. Cloud routers are increasingly being adapted into SDN solutions courtesy of their flexibility, scalability and simple management \cite{b25}. These cloud instances charge users for network traffic egress\cite{b3, b6, b7, b8}, which penalizes paths that take more hops. Traditional SD-WAN solutions overlook this aspect, resulting in increased operational expenses, which could be undesirable from a consumer's perpective\cite{b4}. Consequently, there is a pressing need for a more comprehensive approach to SD-WAN control plane design, one that takes into account the unique requirements of modern enterprises.

To address these challenges, we propose WirePlanner, a novel SD-WAN solution that integrates an advanced path discovery algorithm and the WireGuard VPN technology\cite{b12} for secure and efficient data transmission. WirePlanner optimizes both latency and cost, offering a significant improvement over conventional SD-WAN implementations.

The cornerstone of WirePlanner is its innovative path discovery algorithm, which considers two payment methods: Pay-As-You-Go, where users pay for a fixed amount of bandwidth over a specified period, and Pay-For-Data-Transfer, where users pay for the size of the data transmitted. By incorporating a user-defined monetary cap and an overlay topology of edge routers, WirePlanner calculates a path between nodes that minimizes latency while remaining within budget constraints, and also returns the billing configurations for each edge router along the path.

In addition to its path discovery capabilities, WirePlanner leverages WireGuard VPN technology to establish secure tunnels along the calculated path. WireGuard is a lightweight, high-performance VPN protocol that offers significant advantages over IPSec, including faster setup times, reduced latency, and increased throughput\cite{b9, b12}. By integrating WireGuard into the SD-WAN solution, WirePlanner ensures secure and efficient data transmission, even in the most demanding enterprise environments.

This paper will provide a comprehensive analysis of the WirePlanner control plane, detailing its design, implementation, and performance benefits. We will discuss the challenges of current SD-WAN solutions and demonstrate how WirePlanner addresses these issues by incorporating a novel path discovery algorithm and the WireGuard VPN technology. By evaluating WirePlanner in a simulated environment with different underlay topologies and budget constraints, we aim to showcase its potential to deliver secure, cost-efficient and low-latency networking for modern enterprises.

\section{Background and Motivation}

\subsection{SD-WAN}

Software-Defined Wide Area Networking (SD-WAN) has emerged as a revolutionary technology that is transforming the way enterprises manage and optimize their wide-area networks. By leveraging the principles of software-defined networking (SDN), SD-WAN offers a host of benefits that address the evolving needs of modern organizations, ranging from improved network performance to increased cost-efficiency\cite{b1, b17, b19, b20}. However, despite its numerous advantages, we find that SD-WAN faces several drawbacks in the areas of path discovery and VPN technology integration.

\begin{figure}[htbp]
\centerline{\includegraphics[width=80mm]{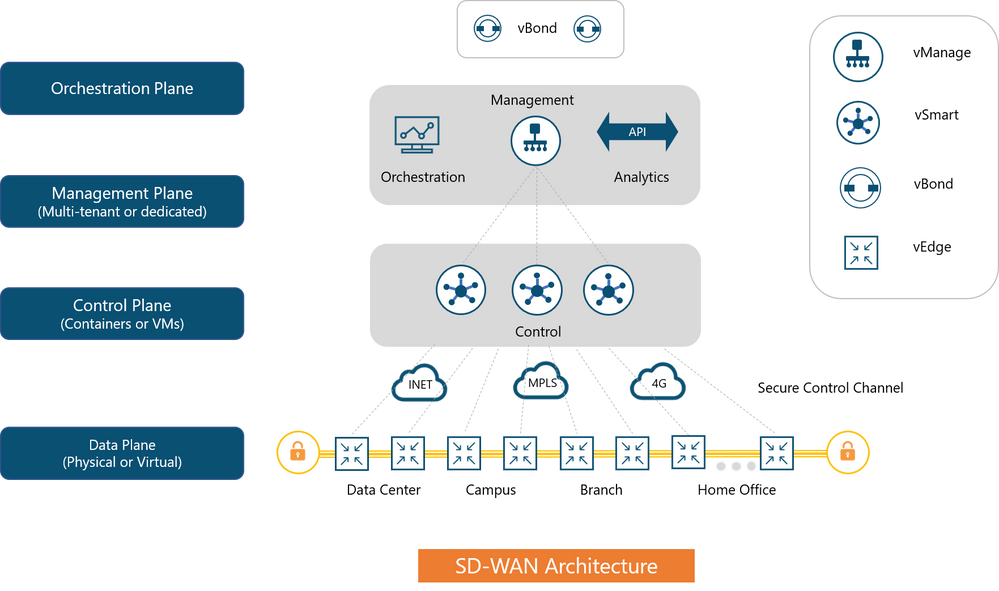}}
\vspace{\baselineskip}
\caption{The logical architecture of a Cisco SD-WAN Solution \cite{b27}}
\label{sdwan}
\end{figure}

The SD-WAN architecture typically comprises four components, from bottom up being the data plane, the control plane, the management plane and the orchestration plane (Fig. \ref{sdwan}) \cite{b1, b2, b18, b21}. The data plane is responsible for transporting data packets between the different locations in the SD-WAN network. It includes the physical and virtual network devices, such as edge routers, switches, and appliances, that forward data packets between sites. The control plane is responsible for managing the data plane and providing network intelligence to optimize network performance. It includes the SD-WAN controllers that provide centralized management and control of the SD-WAN network. Orchestration is the process of automating and managing the deployment and configuration of network services and infrastructure, and involves the use of automation tools and SDN technologies to manage and configure the network dynamically\cite{b1, b18}.

One of the main challenges in SD-WAN architecture lies in the path discovery process. Traditional SD-WAN solutions often focus solely on latency or cost when determining the optimal route between nodes, neglecting the need for a balanced approach that considers both factors. This can lead to suboptimal network performance, increased operational costs, and diminished Quality of Service (QoS). Consequently, there is a growing demand for SD-WAN solutions that can intelligently optimize both latency and cost in path discovery, delivering enhanced performance and cost management.

Another challenge faced by popular SD-WAN implementations is the limitations of the VPN technologies employed. Many solutions, such as Cisco SD-WAN\cite{b14}, Versa SD-WAN\cite{b15} and Meraki SD-WAN\cite{b16} utilize Internet Protocol Security (IPSec) or VPN protocols based on IPSec to establish secure connections between edge routers. While IPSec offers robust encryption, it also introduces long setup times, due to the burdensome Internet Key Exchange (IKE)\cite{b11}, and reduced throughput\cite{b9}. These limitations can negatively impact the overall performance and scalability of SD-WAN deployments, hindering their adoption in various enterprise contexts.

In conclusion, while SD-WAN has emerged as a promising networking paradigm for modern enterprises, it has limitations in path discovery and VPN technology integration. Addressing these challenges requires innovative solutions that consider both cost and latency when determining the optimal route between nodes, as well as the adoption of more efficient VPN technologies. By overcoming these limitations, future SD-WAN implementations can deliver improved performance, cost-efficiency, and security, meeting the evolving demands of today's digital landscape.

\subsection{WireGuard}

WireGuard\cite{b12} is a highly efficient Virtual Private Network (VPN) protocol designed to provide secure and fast communication between devices over the Internet. Developed by Jason A. Donenfeld, WireGuard has gained widespread recognition for its performance, simplicity, and security features.

WireGuard operates by creating encrypted and authenticated tunnels between devices. It leverages modern cryptographic primitives, such as Curve25519 for key exchange, ChaCha20 for symmetric encryption, and Poly1305 for message authentication\cite{b9}\cite{b13}. These choices offer strong security while maintaining minimal overhead, contributing to WireGuard's lightweight design.

One of the defining features of WireGuard is its simplicity. The codebase consists of only a few thousand lines, in stark contrast to other VPN protocols like OpenVPN\cite{b10} or IPsec\cite{b11}, which have tens or even hundreds of thousands of lines of code. This compact size not only makes WireGuard easier to audit for security vulnerabilities but also reduces the attack surface, contributing to a more robust security posture\cite{b9}.

In terms of performance, WireGuard outperforms existing VPN solutions\cite{b22}, thanks to its use of modern cryptographic algorithms and streamlined design. It boasts reduced latency and lower resource consumption\cite{b9}, making it suitable for a wide range of applications, from mobile devices to data centers.

Furthermore, WireGuard's unique "cryptokey routing" mechanism simplifies the configuration process. Instead of relying on complex routing tables and firewall rules, WireGuard associates public keys with allowed IP addresses\cite{b5, b9}. This approach not only eases deployment and management but also enhances privacy, as IP addresses are kept confidential within the encrypted tunnel.

To summarize, WireGuard offers enhanced performance by leveraging modern cryptographic algorithms and efficient design, improved security by using a smaller and more easily audited codebase, and simplified configuration by utilizing "cryptokey routing" and binding public keys with allowed IP addresses.

\subsection{Router Billing}

Cloud routers, also known as virtual routers, represent a significant shift in wide-area network management by introducing the concept of routing as a service. By moving the routing functionality to the cloud, organizations can benefit from increased flexibility, scalability, and cost-efficiency\cite{b25}. Cloud routers enable seamless integration of on-premises and cloud-based resources, simplifying network administration and reducing the need for physical hardware. Furthermore, cloud routers are easily provisioned, managed, and decommissioned according to changing network demands, ensuring a dynamic and adaptable networking infrastructure that aligns with the evolving requirements of modern enterprises\cite{b23, b24}.

Prior work has examined the cloud egress pricing for instances in different cloud providers\cite{b4} with the following results: users are only charged for traffic leaving a cloud region, inter-cloud transfers charge users at a fixed rate, the price of intra-cloud transfers is reliant on geographical distance, and in a bulk data transfer scenario, network egress prices outweigh VM prices by a large margin\cite{b6, b7, b8}. Notably, in our setting, the cloud instances simulate edge routers in an SD-WAN, hence they would always be present regardless of workload, so this cost is neglected in our design.

However, this work only identified a payment method where users are charged based solely on the volume of data transmitted. Alibaba Cloud provides two additional billing methods, namely Subscription and Pay-As-You-Go (PAYG). Table \ref{bill} shows the billing methods and the price rates for network traffic of an ECS instance situated in Singapore\cite{b3}.

\begin{table}[htbp]
\caption{Billing Methods and Price}
\begin{center}
\begin{tabular}{| m{3cm} | m{3cm} | m{1.5cm} |}
\hline
\textbf{Billing Method} & \textbf{Unit} & \textbf{Price}\\
\hline
Subscription, Tiered Pay-By-Bandwidth & 5 Mbps & \$17.0/Month \\
\hline
Subscription, Tiered Pay-By-Bandwidth & 6 Mbps or Greater, at Price per Mbps & \$11.8/Month \\
\hline
Pay-As-You-Go, Tiered Pay-By-Bandwidth & 5 Mbps & \$0.0296/Hour \\
\hline
Pay-As-You-Go, Tiered Pay-By-Bandwidth & 6 Mbps or Greater, at Price per Mbps & \$0.021/Hour \\
\hline
Pay-By-Data-Transfer & 1GB & \$0.081/Hour \\
\hline
\end{tabular}
\label{bill}
\end{center}
\end{table}

When designing our path discovery algorithm, multiple billing methods are taken into consideration, since the performance and cost of each differ depending on circumstance.

\section{WirePlanner Design}
This section introduces the design overview of WirePlanner and explains in detail how WirePlanner discovers a path in the edge router network that optimizes both cost and latency, and how it configures a WireGuard tunnel along this path for secure yet lightweight encryption.

\subsection{Overview}
WirePlanner comprises two components, namely \textbf{WireBrain}, and \textbf{WireMuscle}. WireBrain is a control layer component that handles route discovery and billing configuration; WireMuscle operates at the data layer is responsible for the configuration of WireGuard tunnels along an underlay path.

\subsection{WireBrain}
\label{wirebrain}

The purpose of WireBrain is to find an underlay path in the data plane that introduces the least latency possible, while ensuring that the cost of transferring data does not exceed the user-defined budget.

\noindent\textbf{Abstraction of the Data Layer}. 
We can model the data plane of the SD-WAN using an edge weighted directed graph. Each vertex in the graph represents an instance in the cloud or an edge router that can be used to transmit data, and each edge represents a physical connection between two instances. Each physical connection has two weights, \textit{t} and \textit{c}, attributed to it, which represent the latency and cost of transmitting data over this path respectively. 

\begin{figure}[htbp]
\centerline{\includegraphics[width=80mm]{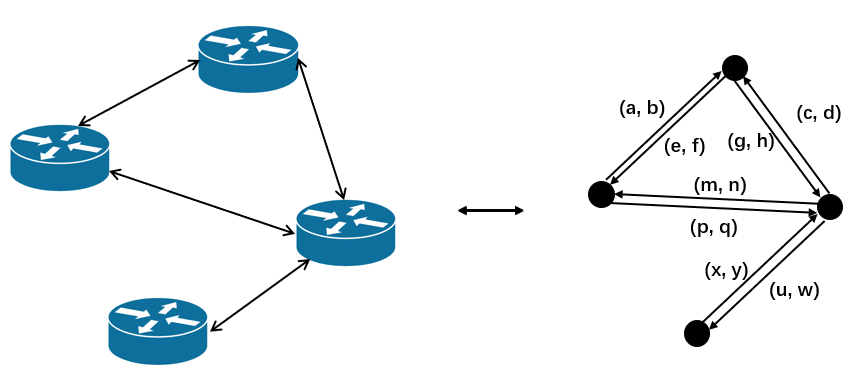}}
\vspace{\baselineskip}
\caption{Example of data layer abstraction}
\label{top}
\end{figure}

\noindent\textbf{Edge Latency}. The latency of a physical connection can be calculated using the equation:
\begin{equation}
    t = t_{prop} + t_{tran}
\end{equation}
WireBrain uses the \textit{ping}\cite{b26} command to obtain the Round-Trip-Time(RTT) between two instances and calculates $t_{prop}$ by dividing the RTT by 2. Given the size of the data to be transferred $d$ and the egress bandwidth of the source instance $b$, $t_{tran}$ can be calculated using $t_{tran} = {d \over b}$.

\noindent\textbf{Edge Cost}. 
Cloud providers charge users for egress traffic leaving an instance and allow users to choose from different payment methods. We identified three types of pricing in Section 2: Subscription(Tiered Pay-By-Bandwidth), Pay-As-You-Go(Tiered Pay-By-Bandwidth) and Pay-By-Data-Transfer respectively. It is impractical to purchase a monthly subscription when performing a singular bulk data transfer that should not take more than a few hours, so WireBrain only considers the latter two methods.

Pay-For-Data-Transfer(PFDT) charges users for the amount of data being transferred, regardless of egress bandwidth. Hence, the cost for PFDT is the product of the data size and the price rate, which are both provided to WireBrain as inputs.

\begin{figure}[htbp]
\centerline{\includegraphics[width=\linewidth]{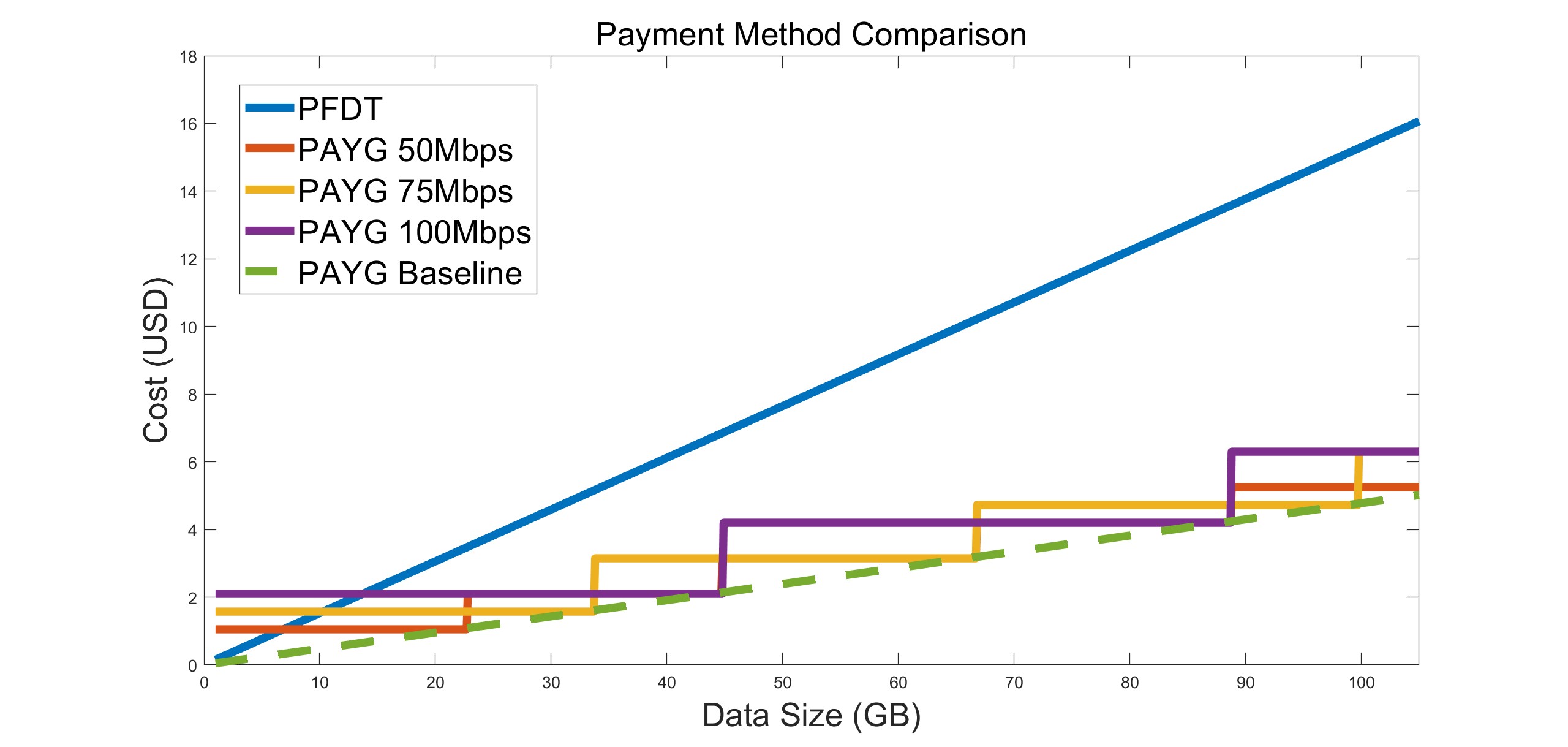}}
\vspace{\baselineskip}
\caption{Edge cost comparison}
\label{payment}
\end{figure}

Pay-As-You-Go(PAYG) charges users per bandwidth per time, regardless of the user's bandwidth utilization. The billed time is rounded up to integer hours, and the cost per hour is approximately proportionate to the amount of bandwidth being purchased\cite{b3}. Therefore, given the billing rate, the purchased bandwidth and the purchased duration, we can calculate the cost by multiplying them together. Notably, when performing a single bulk data transfer, bandwidth and time are interdependent. Figure \ref{payment} shows the relationship between egress cost and data size of an Alibaba ECS instance situated in Hong Kong\cite{b3}.

\noindent\textbf{Path Discovery Algorithm}. We define the parameters relevant to WireBrain in Table \ref{symbols}.

\begin{table}[htbp]
\caption{WireBrain Symbols}
\begin{center}
\begin{tabular}{m{3cm}  m{4cm}}
\hline
\multicolumn{2}{ c }{Input} \\
\hline
$N \in \mathbb{Z}_{+}$ & Number of Instances \\
$G \in \{0,1\}^{N\times N}$ & Graph Adjacency Matrix\\
$B \in \mathbb{R}_{+}^{N}$ & Maximum Egress Bandwidth \\
$s \in \mathbb{Z}$ & Source \\
$d \in \mathbb{Z}$ & Destination \\
$D \in \mathbb{R}_{+}$ & Data Size \\
$Budget \in \mathbb{R}_{+}$ & Budget \\
$PAYG\in \mathbb{R}_{+}^{N}$ & Pay-As-You-Go Price Rate\\
$PFDT\in \mathbb{R}_{+}^{N}$ & Pay-For-Data-Transfer Price Rate \\
$L\in \mathbb{R}_{+}^{N\times N}$ & Latency \\
$I \in \mathbb{Z}_{+}$ & Max Number of Iterations \\
\hline
\hline
\multicolumn{2}{ c }{Output} \\
\hline
$P\in \mathbb{Z}^{N}$ & Path \\
$B\in \mathbb{R}_{+}^{N}$ & Configured Bandwidth \\
$M \in \{0, 1, 2\}^{N}$ & Billing Method (0 for None, 1 for PAYG, 2 for PFDT) \\
\hline
\end{tabular}
\label{symbols}
\end{center}
\end{table}

The design of our algorithm is based on two key insights. a) Regardless of circumstance, PFDT always yields lower or tantamount latency compared to PAYG. b) When the egress bandwidth for PAYG is fixed, which billing method yields the cheapest price is dependent on data size. We identify a data size threshold $D_{thresh}$ that can be calculated given the billing rates of PAYG $k_1$ and PFDT $k_2$ and the egress bandwidth $B$:

\begin{equation}
    k_1 * B = k_2 * D_{thresh}; \ \ D_{thresh} = {k_1*B\over k_2}
\end{equation}

We propose a modified version of the classic Dijkstra Algorithm where each edge has two weights instead of one, called A and B respectively. The objective of this algorithm is to find a path from the source to the destination along which the sum of the A weights does not surpass a given maximum value and the sum of the B weights is minimal.

\begin{algorithm}[htbp]
    \algsetup{linenosize=\tiny} \scriptsize
	\caption{WireBrain\_Search} 
	\label{alg1} 
    \renewcommand{\algorithmicrequire}{\textbf{Input:}}
    \renewcommand{\algorithmicensure}{\textbf{Output:}}
	\begin{algorithmic}[1]
		\REQUIRE $(G, s, d, A, B, A_{max}, N)$
		\ENSURE $P$ - Path 
		\STATE $Q \gets \{\}$
        \STATE $prev \gets [-1;N]$
        \STATE $MinB \gets [+\infty;N]$
        \STATE $MinB[Src] \gets 0$
        \STATE $Q.add((Src, 0, 0))$
        \WHILE{$Q$ is not empty}
            \STATE $node, currA, currB \gets min_{b}(node, a, b) \in Q$
            \STATE $Q.remove((node, currA, currB))$
            \IF{$node = Dst$}
                \RETURN $prev$
            \ENDIF
            \FOR{each edge $(node, nxt) \in G$}
                \STATE $newA \gets currA + A_{node, nxt}$
                \STATE $newB \gets currB + B_{node, nxt}$
                \IF{($newA \leq A_{max}$ and $newB < MinB[nxt]$)}
                    \STATE $MinB[nxt] \gets newB$
                    \STATE $prev[nxt] \gets node$
                    \STATE $Q.add((nxt, newA, newB))$
                \ENDIF
            \ENDFOR
        \ENDWHILE
        \RETURN $None$
	\end{algorithmic} 
\end{algorithm}

The algorithm for path discovery initially configures each node to send data at maximum bandwidth, and their billing methods are determined based on the comparison between data size and their $D_{thresh}$. The cost and latency of each edge can be calculated using each node's bandwidth and billing method, which correspond to weights A and B in the WireBrain\_Search Algorithm. If the WireBrain\_Search Algorithm can produce a valid result with the initial configuration, then the results can be immediately returned; failure of the algorithm indicates the initial billing configuration being too expensive. The measure taken to reduce cost is to halve the bandwidth of each node, use it to calculate a new $D_{thresh}$ and choose a billing method based on that. This process is repeated a total of $I$ times, with each iteration modifying the egress bandwidth using a binary approach. The algorithm returns the last successful path given by WireBrain\_Search and its corresponding billing configuration, and fails if after $I$ rounds of iteration WireBrain\_Search is still unable to produce a result, indicating an insufficient budget.

\subsection{WireMuscle}

For enterprises to safely transfer sensitive data between edge-devices, utilization of data encryption technologies is essential\cite{b18}, and among these technologies, Virtual Private Networks(VPNs) are the most widely adapted. Compared to its competition, WireGuard stands out for offering enhanced performance, simple configuration and minimal startup time by fixing its cryptographic algorithm and implementing a streamlined design\cite{b9}.

After WireBrain returns an underlay path from the source to the destination, WireMuscle establishes WireGuard tunnels along this path by assigning public keys to allowed IP addresses and configuring peer-to-peer connections between adjacent instances. The concatenation of WireGuard tunnels provides a means for data to be encrypted robustly and transferred from end to end with little performance degradation.

\begin{algorithm}[hbp]
    \algsetup{linenosize=\tiny} \scriptsize
	\caption{Path Discovery Algorithm} 
	\label{alg2} 
    \renewcommand{\algorithmicrequire}{\textbf{Input:}}
    \renewcommand{\algorithmicensure}{\textbf{Output:}}
	\begin{algorithmic}[1]
		\REQUIRE $(N, G, B, s, d, D, Budget, PAYG, PFDT, L, I) \ $
		\ENSURE $(P, B, M)$
        \STATE $X, Y \gets \ $ all zero $\mathbb{R}_+^{N\times N}$ matrix
        \STATE $B[i] \gets E[i],\  i=0, 1,..., N-1$
        \STATE $Y_{ij} \gets t_{prop} + t_{tran}$ for each edge
        \FOR{$i$ from $0$ to $N-1$}
            \STATE Calculate $D_{thresh}$ for $node_i$
            \IF{$D<D_{thresh}$}
                \STATE $M[i] \gets 2, \ X_{ij} \gets PFTD[i]*D, \ j\ne i$
            \ELSE
                \STATE $M[i] \gets 1, \ X_{ij} \gets PAYG[i]*E[i]*$time
            \ENDIF
        \ENDFOR
        \STATE $P \gets \ $WireBrain\_Search$ (G, s, d, X, Y, Budget, N)$
        \IF{$P$}
            \RETURN $(P,B,M)$
        \ENDIF
        \STATE $Iter \gets 0, \ k\gets {1\over 2}, \ k_{upper}\gets 1, \ k_{lower}\gets 0$
        \STATE $P_{curr}\gets P,\ B_{curr}\gets B, \ M_{curr}\gets M$
        \STATE $P\gets None,\ B\gets None,\ M\gets None$
        \WHILE{$Iter < I$}
            \STATE $B_{curr}[i] \gets k*E[i],\  i=0, 1,..., N-1$
            \FOR{$i$ from $0$ to $N-1$}
                \STATE Calculate $D_{thresh}$ for $node_i$
                \IF{$D<D_{thresh}$}
                    \STATE $M_{curr}[i] \gets 2, \ X_{ij} \gets PFTD[i]*D, \ j\ne i$
                    \STATE $B_{curr}[i] \gets E[i]$
                \ELSE
                    \STATE $M_{curr}[i] \gets 1,  X_{ij} \gets PAYG[i]*B_{curr}[i]*$time  
                \ENDIF
            \ENDFOR
            \STATE $Y_{ij} \gets t_{prop} + t_{tran}$ for each edge
            \STATE $P \gets \ $WireBrain\_Search$ (G, s, d, X, Y, Budget, N)$
            \IF{$P$}
                \STATE $k \gets {k+k_{upper}\over 2}, \ k_{lower} \gets k$
                \STATE $P\gets P_{curr},\ B\gets B_{curr},\ M\gets M_{curr}$
            \ELSE
                \STATE $k \gets {k+k_{lower}\over 2}, \ k_{upper} \gets k$
            \ENDIF
            \STATE $Iter \gets Iter + 1$
        \ENDWHILE
        \RETURN $(P,B,M)$
	\end{algorithmic} 
\end{algorithm}

\section{Implementation and Evaluation}
We have implemented a prototype of WirePlanner on six ECS instances in Alibaba Cloud, where each instance simulates an edge router in the data plane of an SD-WAN solution. The specific configurations for each cloud instance are shown in Table \ref{table:config}. We deployed WireGuard, OpenVPN and IPSec on the instances situated in Beijing and Shenzhen, and only deployed WireGuard on the remaining four instances. The routing tables of the six instances were reconfigured to simulate different control layer topologies in our experiment. We implemented the algorithms introduced in \ref{wirebrain} using Python 3.

\begin{table}[htbp]
\centering
\caption{Configuration of Cloud Instances Used in Experiment}
\begin{tabular}{|c|c|c|c|}
\hline
Type       & Location       & CPU cores & RAM(GB)   \\ \hline
ECS Ubuntu & Beijing  & 1         & 1         \\ \hline
ECS Ubuntu & Shanghai & 1         & 1         \\ \hline
ECS Ubuntu & Shenzhen & 1         & 1         \\ \hline
ECS Ubuntu & Chengdu  & 1         & 1         \\ \hline
ECS Ubuntu & London      & 2         & 4         \\ \hline
ECS Ubuntu & Virginia    & 1         & 1         \\ \hline
\end{tabular}
\label{table:config}
\end{table}

\subsection{WireGuard Overhead}
\label{exp1}

We first compare the latency, startup time and throughput between WireGuard, IPSec and OpenVPN to demonstrate WireGuard's superiority in performance.

\begin{figure}[htbp]
\centerline{\includegraphics[width=90mm]{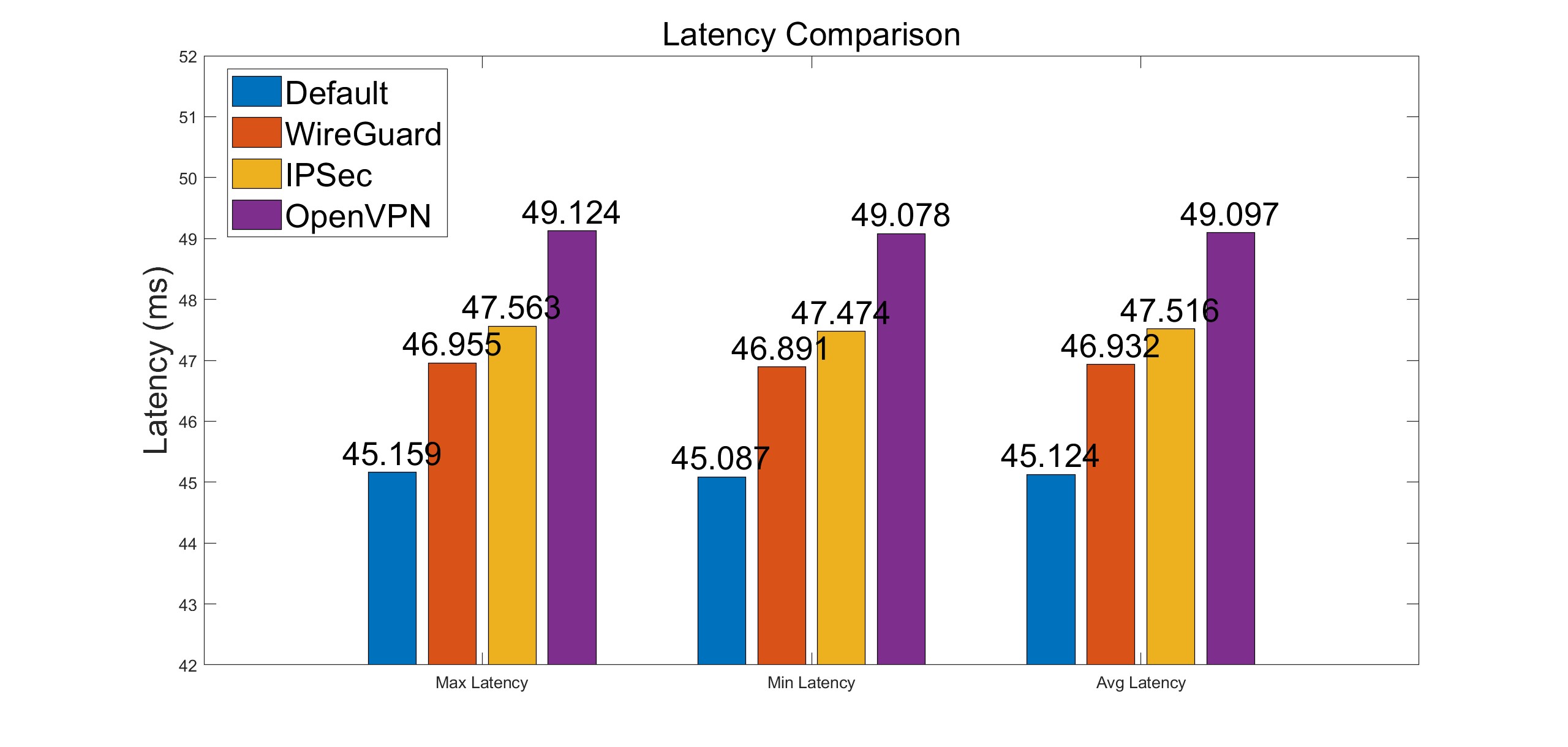}}
\caption{Latency comparison between VPNs}
\vspace{\baselineskip}
\label{latency}
\end{figure}

\begin{figure}[htbp]
\centerline{\includegraphics[width=90mm]{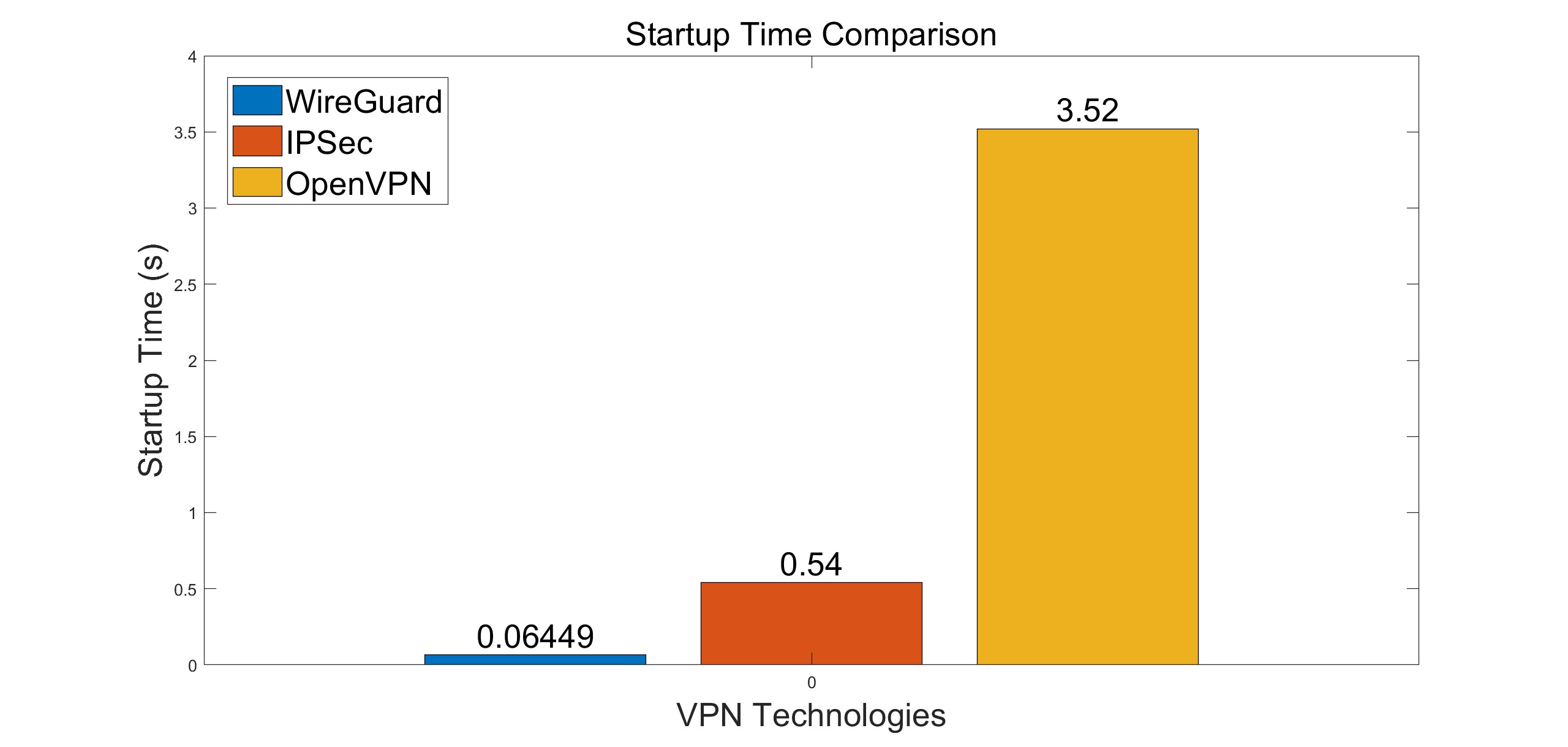}}
\caption{Startup time comparison between VPNs}
\vspace{\baselineskip}
\label{startup}
\end{figure}

\begin{figure}[htbp]
\centerline{\includegraphics[width=90mm]{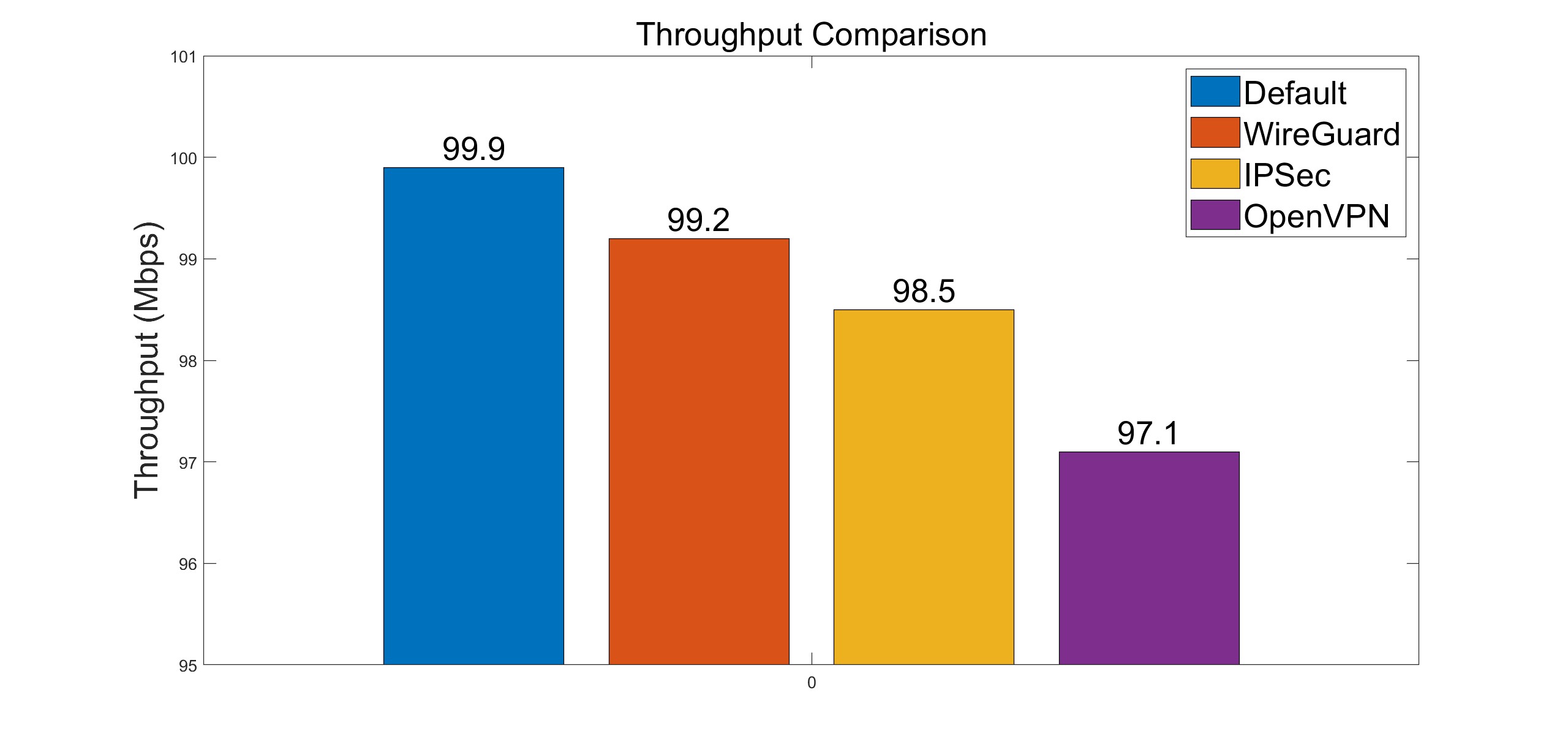}}
\caption{Throughput comparison between VPNs}
\vspace{\baselineskip}
\label{throughput}
\end{figure}

Latency can be obtained by calculating the difference between the time it takes to \textit{ping} between two instances using the VPN tunnel and using the public IP address without configuring the VPN. We measure the time it takes for a tunnel between the Beijing instance and the Shenzhen instance to be fully established to acquire the startup times of each VPN. We also measured the throughput of the three VPN services using \textit{IPERF}; due to the cloud limitations, the egress bandwidth of each instance is limited to 100 Mbps. We average the data over 10 iterations and present them in Figures \ref{latency}, \ref{startup}, and \ref{throughput}.

The results show that though WireGuard is only marginally better than IPSec and OpenVPN in terms of latency and throughput, it has an overwhelming edge over them in startup time.

\subsection{End-to-End Performance}
\label{exp3}
\begin{figure}[htbp]
\centerline{\includegraphics[width=90mm]{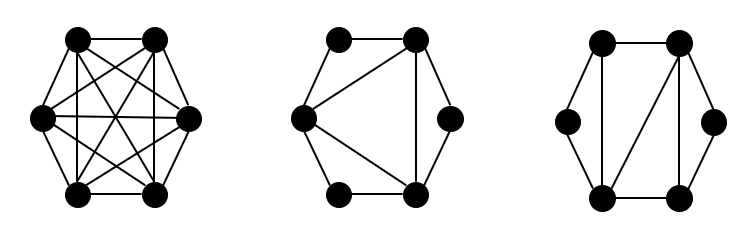}}
\caption{Data plane topologies used in end-to-end evaluation}
\vspace{\baselineskip}
\label{top3}
\end{figure}

This experiment compares the performance of the WirePlanner solution, the WirePlanner solution without WireGuard configured, and a na\"ive routing solution with WireGuard configured. We use three different tolopogies as shown in Figure \ref{top3}, and leverage the price rates and maximum egress bandwidths specified on the Alibaba Cloud website\cite{b3} to find perform the algorithm. We then compared the results from the algorithm with simulations that use real-time throughput and \textit{ping} delay to estimate the actual latency. The three topologies used in this experiment are shown in Figure \ref{top3}, and the source and destination are the leftmost and rightmost nodes in each graph.

Evaluation results are presented in Table \ref{ete}.

\begin{table}[htbp]
\caption{Latency (s)}
\begin{tabular}{|c|c|c|c|}
\hline
           & WirePlanner     & WirePlanner w/o WireGuard & Na\"ive Route  \\ \hline
Top 1 &  82.82  &   82.24      &  248.46        \\ \hline
Top 2 &  165.41 &   164.26       & 248.46         \\ \hline
Top 3 & 165.78 & 164.62         & 248.46         \\ \hline
\multicolumn{4}{ l }{$^{\mathrm{a}}$Data size is 1GB} \\ 
\multicolumn{4}{ l }{$^{\mathrm{b}}$Budget is \$0.5} \\ 
\end{tabular}
\vspace{\baselineskip}
\label{ete}
\end{table}

Simulation results indicate WirePlanner only lacks $0.7\%$ in performance compared to a solution that travels the same route without WireGuard, and consistently outperforms na\"ive routing by at least 30\%.

\section{Conclusion}
In this paper we presented WirePlanner, an SD-WAN solution that discovers a path between two nodes that optimizes both latency and cost, configures the billing method of each node along the path and sets up WireGuard tunnels for secure transmission. WirePlanner uses a combination of WireBrain and WireMuscle for smart routing and robust encryption, and evaluation results indicate that WirePlanner can successfully achieve fast and secure data transmission under budget.

\end{document}